\documentclass{egpubl}
\usepackage{sca2026}
 
\SpecialIssuePaper         

\CGFccby

\usepackage[T1]{fontenc}
\usepackage{dfadobe}  

\usepackage{cite}  
\BibtexOrBiblatex
\electronicVersion
\PrintedOrElectronic

\ifpdf \usepackage[pdftex]{graphicx} \pdfcompresslevel=9
\else \usepackage[dvips]{graphicx} \fi

\usepackage{egweblnk} 

\usepackage[utf8]{inputenc}
\usepackage{amsmath}
\usepackage{yfonts}
\usepackage{color}
\usepackage{bm}
\usepackage{mathtools}
\usepackage{multicol}
\usepackage{xfrac}
\usepackage{makecell}
\usepackage{siunitx}
\usepackage{duckuments}
\usepackage{amssymb}

\usepackage{tabularx}
\usepackage{booktabs}
\usepackage{wrapfig}
\usepackage{booktabs} 

\usepackage{pifont}

\newcommand{\cmark}{\ding{51}}
\newcommand{\xmark}{\ding{55}}

\usepackage[boxed,ruled,linesnumbered]{algorithm2e} 
\SetAlFnt{\small}
\SetAlCapFnt{\small}
\SetAlCapNameFnt{\small}
\SetAlCapHSkip{0pt}

\SetCommentSty{mycommfont}

\newcommand{\ff}{\bm{f}}

\definecolor{SithColor}{rgb}{0.7,0,0} 

\definecolor{GuardianColor}{rgb}{0,0,0.8} 

\definecolor{ConsularColor}{rgb}{0,0.4,0} 

\title[Fluid Control with Localized Spacetime Windows]%
      {Fluid Control with Localized Spacetime Windows}
\author[Y. Chen \& D. I.W. Levin \& T. R. Langlois]
{\parbox{\textwidth}{\centering Yixin Chen$^{1}$\orcid{0000-0001-7547-9587}, David I.\,W. Levin$^{1,2}$\orcid{0000-0001-7079-1934} and 
Timothy R. Langlois$^3$\orcid{0000-0002-5043-8698}}
        \\
{\parbox{\textwidth}{\centering $^1$University of Toronto, Toronto, ON, Canada\\
    $^2$Nvidia, Toronto, ON, Canada\\
    $^3$Adobe, Seattle, WA, USA
       }
}
}

\begin{document}

\teaser{
\includegraphics[width=\linewidth]{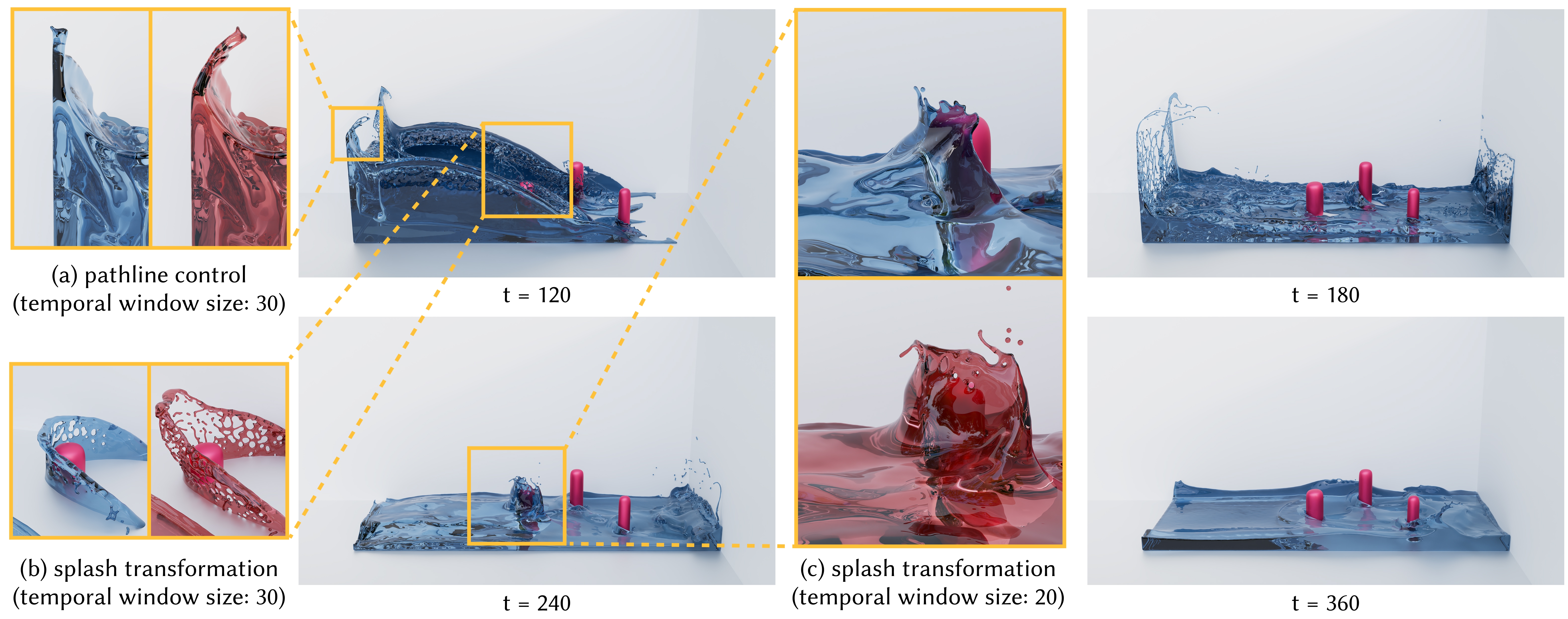}
 \centering
   \caption{\textbf{Large-scale fluid simulation with localized editing and control. Given a free-surface simulation with 1.08 million particles (DOFs of full simulation), we demonstrate that localized editing and spacetime control can be achieved efficiently by optimizing the control forces on a sparse control grid with fewer than 400 grid nodes (DOFs of optimization) within a small temporal window. Our approach greatly reduces the dimensionality of the control problem. For each user-specified edit, we compare the original animation (top/left: blue) with the controlled animation (bottom/right: red), highlighting targeted fluid behaviors such as pathline control (a) and splash transformation (b and c).}}
\label{fig:teaser}
}

\maketitle
\begin{abstract}
   We present a physics-based fluid control method utilizing localized spacetime windows, extending force-based fluid control to substantially larger simulation scales. 
   In many practical editing scenarios, user-specified objectives affect only a small region of an otherwise satisfactory simulation, resulting in optimal control force distributions that are highly sparse in both space and time. 
   However, existing optimization-based fluid control methods typically solve for control forces over the entire spacetime domain, leading to unnecessarily high computational cost and poor scalability.
   Motivated by this observation, we restrict optimization to localized spacetime regions surrounding the edit of interest, significantly reducing the dimensionality of the control problem.
   Within this framework, control forces are parameterized on a coarse "floating" background grid, decoupling control degrees of freedom from simulation resolution and promoting smooth, physically plausible forces. We further analyze spacetime-window selection as a joint spatial-temporal problem. While the full problem can be formulated as a 2D search over spatial and temporal window extents, practical workflows can often leverage user-specified spatial regions and lightweight temporal-window selection strategies to reduce search cost.
   Our method enables a range of intuitive editing tasks, where sparse user inputs can induce coherent motion in surrounding fluid structures.
   We demonstrate the effectiveness and efficiency of our method with various 2D and 3D particle-based free-surface simulation examples.


\end{abstract}  
\section{Introduction}
\label{sec:intro}

\begin{figure*}[!ht]
    \centering
    \includegraphics[width=\textwidth]{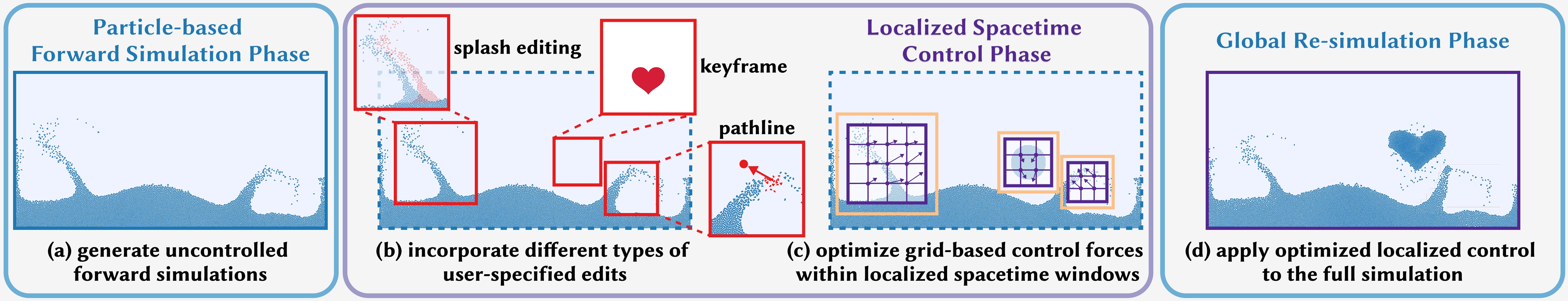}
    \caption{\textbf{Overview of our localized spacetime fluid control framework.} Given a forward simulation (a) and user-specified localized objectives (b), we restrict optimization to selected spacetime regions where control is most effective. Within these regions, control forces are parameterized on a coarse grid and optimized (c), significantly reducing the dimensionality of the problem. The optimized local control is then applied to the full simulation through global re-simulation (d), producing coherent and physically plausible results.}
    \label{fig:control_pipeline}
\end{figure*}

Simulating fluids has long been a crucial topic in computer graphics, with numerous techniques developed to produce visually appealing animations. However, controlling fluid simulations, especially liquids, in a fast and responsive way remains challenging. At the heart of the problem is the need to balance computational efficiency and physical plausibility. Traditional optimization-based control methods yield natural and visually compelling results, yet they often suffer from high computational cost, limiting them to offline control applications. In contrast, optimization-free methods offer relatively good runtime performance, but tend to sacrifice the realism and consistency of fluid behavior. As the demand for interactive, user-in-the-loop fluid control increases in domains such as design tools and visual storytelling, existing methodologies reveal fundamental limitations.

We observe that, in many practical scenarios, users are often satisfied with the overall fluid simulation but seek to make localized refinements, e.g., "the direction of this splash should change a bit", "water shouldn't splash out of the container here", "the fluid should speed up/slow down here". These localized editing objectives suggest a structural property of the underlying optimization problem: although the simulation may be large, the control forces required to achieve a localized edit are often sparse in both space and time. In other words, when the editing objective is localized, the solution (control forces) of a \emph{global} fluid control problem is \emph{localized} around the edit of interest (see Fig.~\ref{fig:global_vs_localized_force}), suggesting that for such edits, a global optimization is unnecessary. This observation motivates our central idea: localizing the inverse control problem itself. Enabling such control, by finding an appropriate localized spacetime window to optimize control forces within, can enhance performance by reducing the size of the optimization problem.

In this paper, we introduce a physics-based framework for fluid control that explicitly exploits this spacetime sparsity. Our approach restricts optimization to localized spacetime regions around user-specified edits, significantly reducing the dimensionality of the control problem. We employ particle-based simulation methods for forward simulations, while representing control forces on a co-located background grid, thereby decoupling the control degrees of freedom from the simulation resolution. More specifically, our high-level technical contributions are:
\begin{itemize}
    \item \textbf{Localized spacetime optimization for free-surface fluid control}, which reformulates fluid control as a localized optimization problem defined only within a bounded spacetime region surrounding the edit. This substantially reduces the dimensionality of the optimization while preserving physically plausible fluid behavior.
    \item \textbf{A coarse background control grid} that decouples control and simulation DOFs. Unlike prior hybrid simulation approaches~\cite{zhu-2005, jiang-2015, stomakhin-2013}, which use grid representations for forward simulation, we instead employ the grid as a control parameterization, where it serves to reduce the dimensionality of the optimization problem and enforce smooth, physically plausible control forces.
    \item \textbf{A practical spacetime-window analysis and selection framework}, based on the observation that effective control is often localized in both space and time. We formulate window selection as a joint spatial-temporal problem, analyze the trade-off between control quality and optimization complexity, and demonstrate practical workflows ranging from joint spacetime search using CMA-ES to user-guided spatial windows with lightweight temporal-window selection.
\end{itemize}
We demonstrate that our method enables efficient and intuitive localized fluid editing across a range of scenarios.

\section{Related Work}
\label{sec:related_work}
\subsection{Liquid Simulation}
Inspired by seminal work~\cite{foster-1996, muller-2003}, the study of detailed, high-performance liquid simulations has emerged as a significant focus in computer graphics research. A comprehensive overview of simulation techniques can be found in \cite{bridson-2015}. While our method could be adapted to various differentiable simulation methods, we utilize Position-Based Fluid (PBF) \cite{macklin-2013} as an example, leveraging its ability to efficiently produce visually plausible results while maintaining stable simulations even with large time steps.

\subsection{Liquid Control}
\label{sec:control_related_work}
Starting from \cite{foster-1997}, fluid control has been an important goal in computer graphics, trying to bridge the gap between physically plausible fluid motion and artistic user intents.

\subsubsection{Optimization-Free Control}
Optimization-free control methods aim to control existing liquid simulations directly without defining and solving expensive optimization problems. Among those, one of the most commonly used strategies is to control liquid motions with user-specified keyframes and skeletons. \cite{mihalef-2004} introduces a control pipeline for breaking waves, allowing animators to define the wave shape at a specific moment using a library of wave profiles. Both \cite{raveendran-2012} and \cite{zhang-2015} utilize a set of 3D meshes as keyframes to guide fluid motion with some physical guidance, such as density constraints, adaptive springs, and velocity adjustments. Inspired by skeletal animation, \cite{zhang-2011} proposes a skeleton-based keyframe control method, which used skeletal structures to enable solid-like liquid motion and shape deformation. Similarly, \cite{lu-2019} introduces a rigging-skinning scheme that enables fluid animations by decoupling control into a rigging phase for low-frequency motion design and a skinning phase for generating plausible flows with adjustable detail without iterative optimization. A recent achievement by \cite{zhou-2024} presents a target-driven fluid simulation method that enhances shape matching by incorporating spatially weighted control, adaptive driving constraints, and density-based incompressibility enforcement. Although these keyframe- and skeleton-based techniques are straightforward, they are often limited by the difficulties of generating accurate liquid keyframes and producing plausible fluid-like behavior.

Beyond user-specified keyframes, some other methods exploit precomputed fluid simulation datasets for controlling and generating new animations. \cite{raveendran-2014} introduces a method for smoothly interpolating between existing liquid animations using a spacetime non-rigid iterative closest point (ICP) algorithm under user guidance, while \cite{manteaux-2016} develops an interactive system for editing precomputed liquid features directly in space and time without requiring re-simulation. Generalized non-reflecting boundary conditions and localized re-simulation techniques~\cite{bojsen-2016} are designed to efficiently update a modified region of a fluid simulation while reducing artifacts at the boundary with the unchanged domain, and the adaptive simulation method~\cite{shah2004extended} exploits spatial locality to dynamically allocate computational resources and improve the efficiency of forward simulation. However, they do not address the inverse control problem considered in our work: given sparse objectives, we optimize force parameters inside a bounded spacetime window and then apply those forces back to the full simulation. The Fluxed Animated Boundary (FAB) method~\cite{stomakhin-2017} controls particle-based fluid simulations by enforcing volumetric flux at boundaries. This technique allows artists to guide fluid behavior using predefined control shapes and flow fields. More recently, a template-based control method is proposed for particle-based simulations \cite{schoentgen-2020}, where precomputed fluid behaviors are transferred to new simulations through global control forces and temporary control particles. Most of these methods provide fast control, but they are potentially limited by precomputed datasets, leading to a limited diversity of achievable motions.

Interactive and sculpting-inspired techniques have also emerged, offering more artist-friendly workflows for users to manipulate and sculpt the liquid.
\cite{stuyck-2016} showcases an interactive, sculpting-inspired approach to fluid animation, allowing artists to directly shape fluid while maintaining physical properties like surface tension and volume preservation using guided re-simulation. While intuitive, such direct manipulation struggles with plausibility, especially for highly dynamic scenarios. \cite{yan-2020} develops an interactive VR-based sketching system for modeling liquid splashes, leveraging a conditional generative adversarial network (cGAN) trained on physical simulations, enabling the fast generation of splash shapes from simple user strokes, though the system primarily targets static splashing shape design and is less suited for controlling dynamic liquid behaviors.

\begin{table}[t]
\centering
\small
\begin{tabular}{c c c c}
\toprule
Method & {Interaction} & {\thead{Control Force \\ Optimization}} & {\thead{Spacetime-localized \\ Optimization}} \\
\midrule
\cite{pan-2013} & {\thead{Geometric \\ Editing}} & \xmark & \xmark  \\
\cite{bojsen-2016} & {\thead{Localized \\ Re-simulation}} & \xmark & \xmark \\
Ours & {\thead{Physics-based \\ Control}} & \cmark & \cmark \\
\bottomrule
\end{tabular}
\caption{
\textbf{Comparison with prior localized simulation and editing methods.}
Unlike previous approaches that localize forward simulation or editing operations, our method localizes the inverse control optimization itself, enabling force-based fluid control within bounded spacetime regions.}
\label{tab:comparison_related_work}
\end{table}

\subsubsection{Optimization-Based Control} 
Formulating fluid control as an optimization problem \cite{treuille-2003} is common, aiming to find the "optimal" (usually meaning as small and smooth as possible) control forces or velocity fields that match user-specified objectives. However, it is computationally expensive to solve a numerical optimization in such a chaotic and high-dimensional system, so prior work has focused on increasing speed. \cite{mcnamara-2004} defines an objective function to measure the difference between the current state of the fluid and the target state, and employs the adjoint method to compute derivatives more efficiently. However, due to computational demands, this pipeline is limited to low-dimensional control forces, leading to overly smooth simulation results. \cite{pan-2017} improves control performance by leveraging spacetime optimization with ADMM and exploiting simulation coherence to significantly reduce computation, though at the cost of introducing high-frequency visual artifacts. Other reduced-order models~\cite{tang-2021, chen-2024}, flow-map-based adjoint solvers~\cite{li2025adjoint} and multi-stage optimization strategies~\cite{kong2025hierarchical} have been proposed to improve control efficiency, unfortunately only supporting keyframe-based smoke control. Notably, \cite{pan-2013} introduces a liquid control framework which allows users to manipulate fluid simulations using keyframe edits, sketches, and small mesh patches in a localized spatial area. By formulating the local changes as a nonlinear geometric optimization problem, this method efficiently propagates edits while preserving volume. Nevertheless, as only geometric constraints are enforced, the resulting animations can sometimes exhibit unnatural transitions when blending with the original simulation sequence. \cite{stuyck-2016-MPC} presents a model predictive controller (MPC) for fluid simulations, which optimizes the control problem with high precision by predicting future states. By using a simplified simulation, the method reduces artifacts and oscillations, resulting in more stable and responsive control, while sacrificing high-frequency fine details of fluid behavior. 

Recent advancements in deep learning have opened new directions for fluid control. \cite{schenck-2018} introduces Smooth Particle Networks (SPNets), a fully differentiable framework integrating fluid dynamics into deep neural networks, allowing learning fluid parameters, performing liquid control and training fluid manipulation policies through end-to-end differentiable optimization. \cite{guan-2022} introduces NeuroFluid, an unsupervised framework that infers fluid state transitions from sequential multiview observations using a particle-driven neural radiance field model. NeuroFluid provides an alternative, learning-based pathway for keyframe-based fluid control. \cite{tao-2024} proposes Neural Implicit Reduced Fluid Simulation (NIRFS), learning reduced latent space dynamics to achieve highly efficient and detailed fluid simulations. They defined the inverse fluid design as an optimization problem and solved it for the optimal initial conditions. Although promising for fast simulations, their ability to generalize across diverse fluid control tasks remains limited. Recent work has explored reinforcement learning-based control for coupled solid–fluid systems, achieving stable control through policy optimization~\cite{chen2025fast}. In contrast, our method targets fluid control, enabling localized editing through optimization.

\section{Background and Motivation}
\begin{figure}[ht]
    \centering
    \includegraphics[width=\columnwidth]{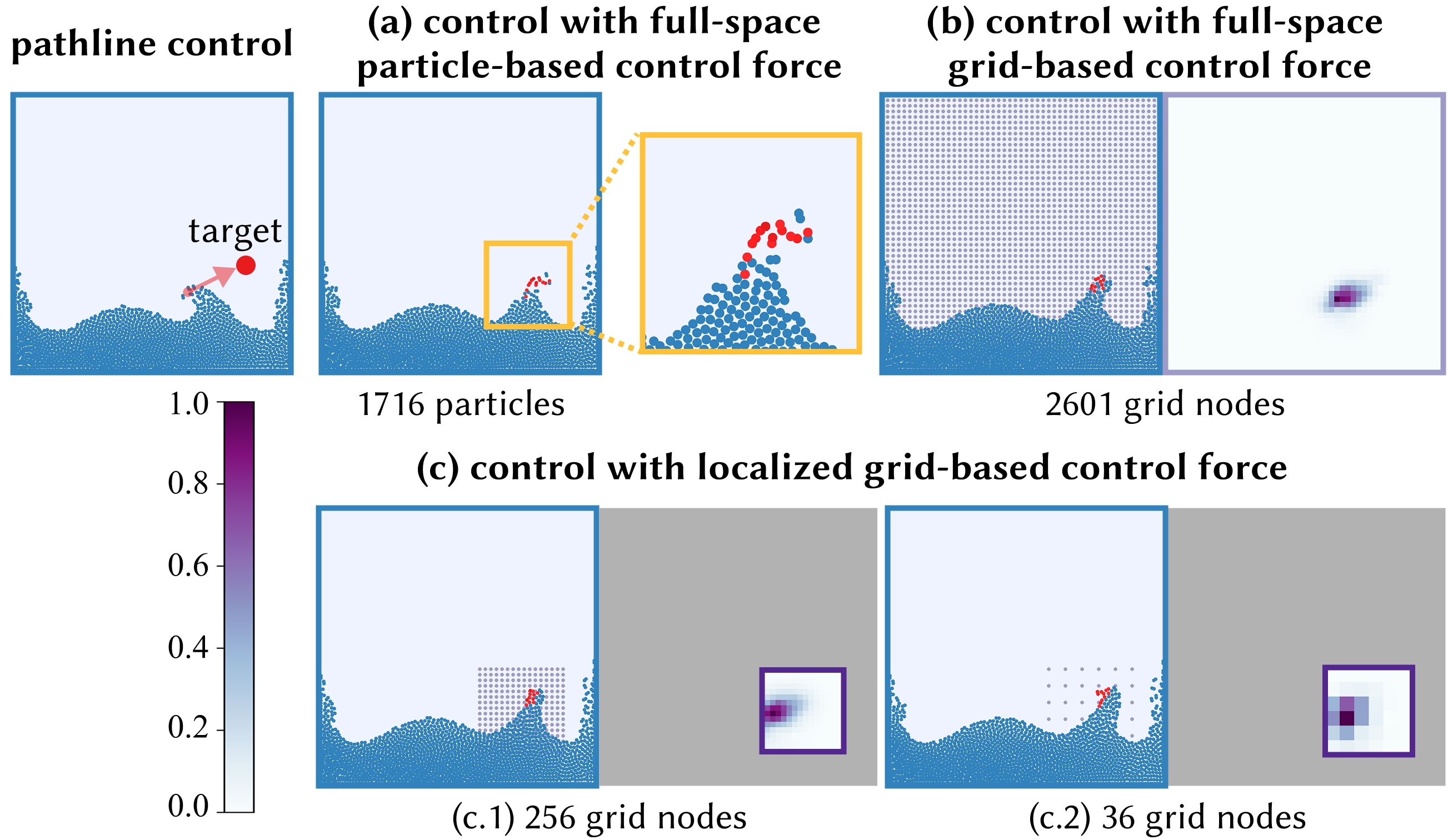}
    \caption{\textbf{Localized control.} Given a requested local edit (the tip of the splash should move), the solution of a global grid-based spacetime optimization (b) is often localized. Restricting the control problem to a local window from the start (c.1) gives a similar solution at a fraction of the cost. Further reducing the localized grid to 36 nodes (c.2) still preserves the desired motion while greatly reducing the optimization cost. Note that the solution when forces are represented on particles (a), instead of a floating grid, results in inconsistencies between the tip and neighboring particles.}
    \label{fig:global_vs_localized_force}
\end{figure}

As stated in \S\ref{sec:control_related_work}, most existing methods for fluid control are often computationally inefficient due to the inherently high-DOF nature of fluids. Several approaches have attempted to reduce DOFs by changing the representation of the control forces~\cite{tang-2021, chen-2024, mcnamara-2004}. In contrast, we take an alternate approach by \emph{reducing the spatial and temporal domain size} of the control problem. This design is motivated by following observations:
\begin{itemize}
    
\item It is often straightforward to set up a simulation to produce the overall desired motion (e.g., water flowing down a rocky slope), but achieving specific local details through manual adjustment of simulation parameters or initial conditions can be painstaking (e.g., modifying the shape or trajectory of a particular splash generated by a rock at a specific moment).

\item Localized editing objectives often induce control solutions that are themselves highly localized in space and time. As a result, solving a global control problem using control degrees of freedom distributed over the entire simulation domain is frequently unnecessary. This observation is illustrated in Fig.~\ref{fig:global_vs_localized_force}.

\end{itemize}
We emphasize that similar observations regarding spatial locality have appeared in previous simulation literature. Multiple methods have used localized regions in various ways for forward simulation to selectively add detail where necessary/desired~\cite{losasso-2004, english-2013, nielsen-2011, bojsen-2016, nielsen-2017, popinet2003, chentanez2011}. Previous work has also used spacetime windows for character animation control~\cite{Cohen:1992:spacetime}. However, we are unaware of the concept being used in fluid control methods. Therefore, we propose a localized control framework (Fig.~\ref{fig:control_pipeline}) that allows for spatially and temporally targeted adjustments, enabling efficient refinement of fluid behavior without altering the overall simulation behavior. 

\subsection{Forward Simulation with Position-Based Fluid (PBF)}
\label{sec:pbf_overview}
Our control framework is designed to be compatible with differentiable fluid simulators. Here we adopt PBF framework~\cite{macklin-2013} for forward free-surface flow simulations, a particle-based method. Unlike traditional smoothed Particle Hydrodynamics (SPH) methods~\cite{koschier-2022}, PBF enforces incompressibility by formulating and solving density constraints, which allows larger time steps and relaxes the neighborhood requirements. Additionally, the method is inherently parallelizable and differentiable, making it particularly well suited for our optimization-based control pipeline. A full description of PBF can be found in~\cite{macklin-2013}. 

\subsection{Connections to Our Control Pipeline} 
A key property of PBF is that particle updates are spatially local: the state of each particle depends only on its neighbors within a finite kernel support radius $h$. More specifically, the kernel function $W(d, h)$ is 0 when the distance between particles $d > h$. This locality implies that changes in particle states propagate gradually over space and time, and can be effectively captured within a bounded spacetime region. This directly influences our choice of spacetime window in~\S\ref{sec:spacetime_window}. By restricting control to such regions, we can effectively influence the desired behavior while significantly reducing the dimensionality of the optimization problem.

\section{Localized Spacetime Control}
\label{sec:method}

We formulate the control as a typical spacetime optimization problem, where plausible control forces are sought to satisfy user-defined objectives. 


\begin{algorithm}[t]
\DontPrintSemicolon
\caption{Localized Spacetime Fluid Control Pipeline}
\label{alg:overall_algorithm}

\KwIn{Initial particle state $\mathbf{x}_0$, simulation parameters $\Theta$, user edit $\mathcal{E}$, user-specified spatial control region $\Omega$}
\KwOut{Controlled fluid animation $\{\mathbf{x}_t^\star\}_{t=0}^{T_{\mathrm{sim}}}$}

Run an uncontrolled forward simulation to obtain the baseline trajectory
$\{\mathbf{x}_{t,\mathrm{orig}}\}_{t=0}^{T_{\mathrm{sim}}}$ \;

Construct the editing objective $\phi_{\mathrm{editing}}$ from the user edit $\mathcal{E}$ \;

Construct the buffer region $\mathcal{B}$ surrounding $\Omega$ \;

Select the temporal control window $[t_s,t_e]$ using either a heuristic rule or CMA-ES (Alg.~\ref{alg:cma-es-alg}) \;

Construct a floating background control grid $\mathcal{G}$ over $\Omega \times [t_s,t_e]$ \;

Initialize localized grid-based control forces
$\mathbf{f} \leftarrow \mathbf{0}$ \;

\While{not converged}{
    Apply grid-based control forces $\mathbf{f}$ to particles inside $\Omega$ through interpolation \;

    Run differentiable localized forward simulation over $[t_s,t_e]$ to obtain
    $\{\mathbf{x}_t\}_{t=t_s}^{t_e}$ \;

    Evaluate the objective
    $\Phi(\mathbf{f}) = \phi_{\mathrm{editing}} + \phi_{\mathrm{force}} + \phi_{\mathrm{buffer}}$ \;

    Compute gradients $\nabla_{\mathbf{f}}\Phi$ by Autodiff backpropagation through the simulation \;

    Update $\mathbf{f}$ using L-BFGS \;
}

Re-simulate the full animation from $\mathbf{x}_0$ while applying the optimized localized control forces $\mathbf{f}^\star$ within $\Omega \times [t_s,t_e]$ \;

\Return{$\{\mathbf{x}_t^\star\}_{t=0}^{T_{\mathrm{sim}}}$} \;
\end{algorithm}

\subsection{Control Force Representation}\label{sec:floating_grid}
Since the forward simulation uses a Lagrangian representation to track the state of individual particles, it is conceptually straightforward to apply control forces on each particle directly. However, such per-particle control not only incurs substantial computational cost but also tends to produce undesirable behavior, such as excessively high-frequency forces or non-physical motion, particularly when fine-grained control is required in localized regions of space and time~(Fig.~\ref{fig:global_vs_localized_force}(a)).
\begin{wrapfigure}{r}{0.55\columnwidth}
    \hspace{-5pt}
    \includegraphics[width=0.55\columnwidth]{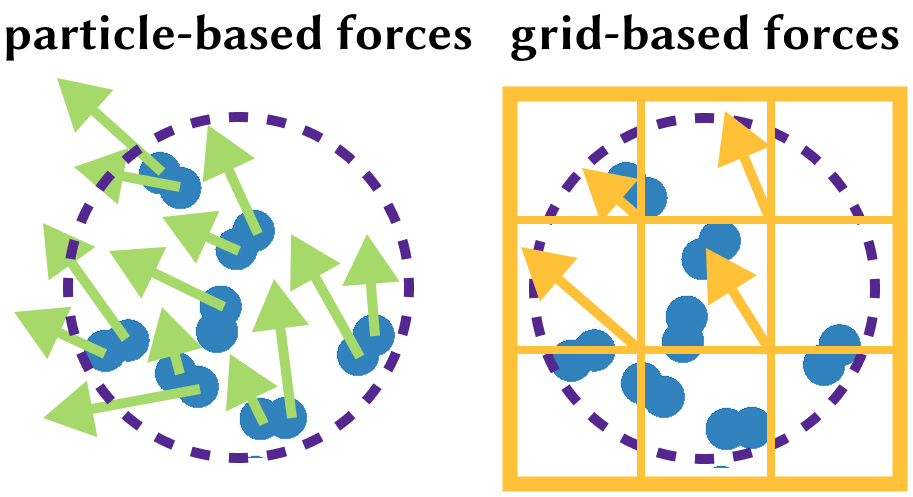}
    \vspace{-5pt}
    \label{fig:particle_grid_visualization}
\end{wrapfigure}
To mitigate these issues, we instead adopt a grid-based control force representation over a background Eulerian grid co-located with the simulation domain. This design is inspired by hybrid particle–grid representations commonly used in simulation methods~\cite{zhu-2005, jiang-2015, stomakhin-2013}, where physical quantities are transferred between particles and grid nodes. However, unlike these approaches, the grid in our framework is not used to evolve physical states, but instead serves as a low-dimensional parameterization of control forces for optimization. 

At each time step, control forces defined on the Eulerian grid are smoothly distributed to nearby particles using a Gaussian kernel, though other interpolation kernel functions can be used without significantly affecting performance. The control force for a particle located at position $\mathbf{x}^p$ is the weighted sum of the control forces from the neighboring grid nodes:
\begin{equation}
\mathbf{f}^{p} = \sum_{i} w_i \cdot \mathbf{f}^{g_i}.
\end{equation}
where $w_i = \exp\left(-\sfrac{d_i^2}{2\alpha^2}\right)$, $\alpha = 0.5 h$, with $h$ being the grid spacing, and $d_i$ is the distance between the particle and the neighboring grid node $i$. The choice of kernel for weighting the control forces is not unique, and alternatives such as the grid-to-particle transfer kernels commonly used in hybrid methods (e.g., FLIP~\cite{zhu-2005}, PIC~\cite{harlow-1962particle} or APIC~\cite{jiang-2015affine}) could also be considered.

This decoupling allows the background grid to be flexible in resolution, adapting seamlessly to different spatial scales without imposing constraints on the simulation. It reduces the dimensionality of the control problem while promoting smooth, spatially coherent force distributions (Fig.~\ref{fig:global_vs_localized_force}(c)), and improves numerical stability during optimization, mitigating the artifacts observed in both per-particle control.


\begin{figure*}[!ht]
    \centering
    \includegraphics[width=\textwidth]{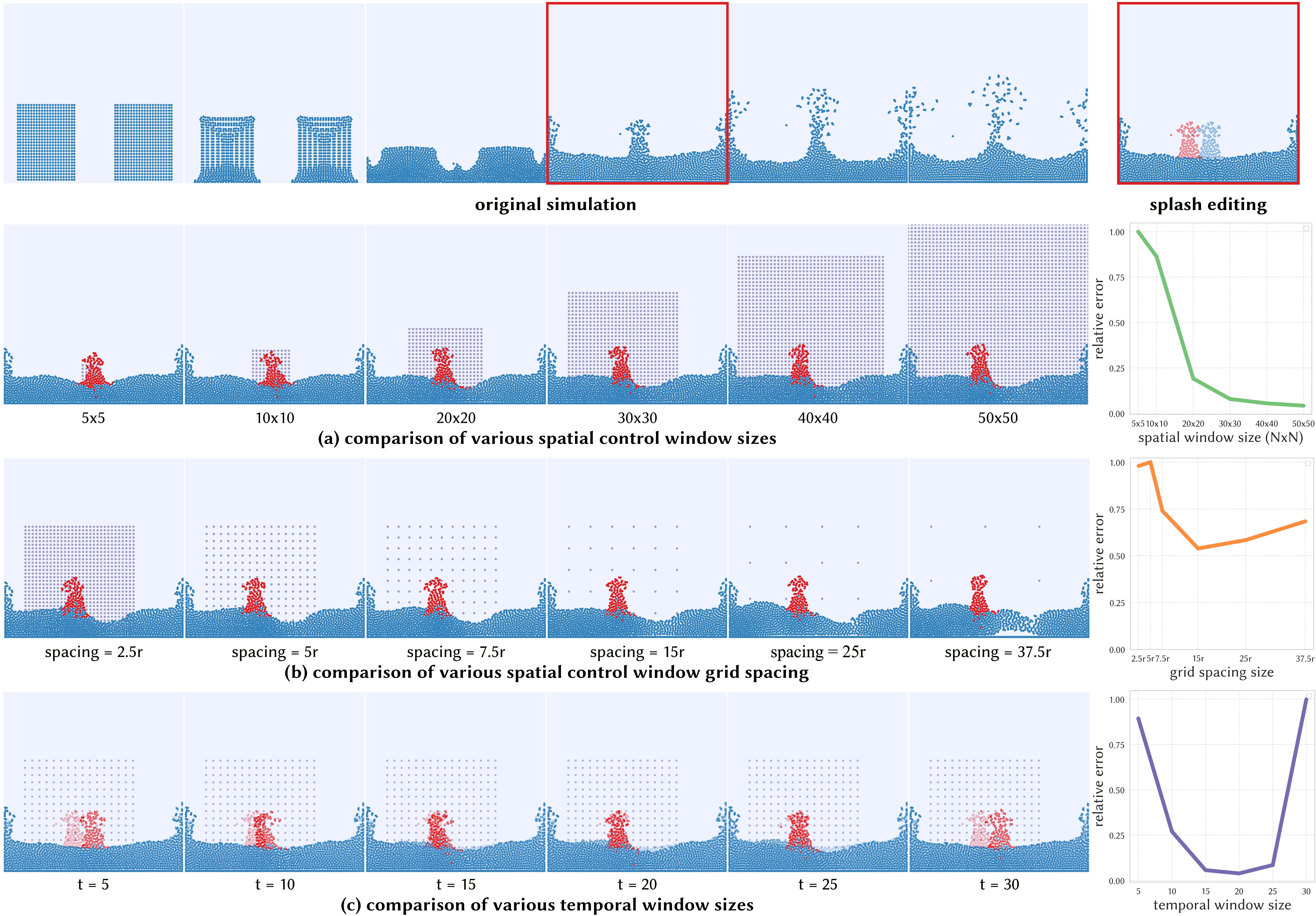}
    \caption{\textbf{Evaluation of 2D localized splash editing under varying control parameters.}
    Top row: The original simulation (left, blue) and a user-specified splash translation at a time in the middle of the simulation (right, red).
    (a) Comparison of different spatial control window sizes. Larger windows match the target better, while smaller windows (e.g., 5×5) struggle to achieve the desired effect. We can observe that once the spatial control grid is big enough to cover the region of interest, the control
    accuracy tends to converge to a satisfactory level. Objective function values are shown in the plots on the right.
    (b) Comparison of different spatial control grid spacings. Extremely fine grids will introduce high-frequency artifacts, while overly coarse grids lead to inconsistency among neighboring particles. The plot on the right illustrates the sweet spot that balances the trade-off.
    (c) Comparison of various temporal window sizes. Very short windows may cause impulsive motion or optimization failure, while overly long windows reduce control efficiency and make the optimization problem challenging. Note that, in theory, larger temporal windows define a superset of the control space and therefore cannot worsen the global optimum. However, they also introduce a higher-dimensional optimization problem, making effective optimization more difficult under a fixed computational budget.}
    \label{fig:splash_editing_comparison}
\end{figure*}

\subsection{Optimization-Based Control Problem}
\label{sec:optimization-based_control_formulation}
We formulate the objective function as a weighted sum of three components: 1) an editing constraint $\phi_{\text{editing}}$, 2) force-related constraints $\phi_{\text{force}}$, regularizing both the magnitude and spatiotemporal smoothness of the control forces, and 3) a buffer-region constraint $\phi_{\text{buffer}}$. The first two components are commonly used in fluid control, while the buffer constraint is introduced to adapt control to localized regions (see~\S\ref{sec:buffer_constraint}). Our goal is to find control forces that minimize this objective:

\begin{equation}
\ff^* = \arg\min_{\ff} (\phi _{\text{editing}} + \phi _{\text{force}} + \phi _{\text{buffer}}).
\label{equ:objective_func}
\end{equation}

\begin{figure*}[!ht]
    \centering
    \includegraphics[width=\textwidth]{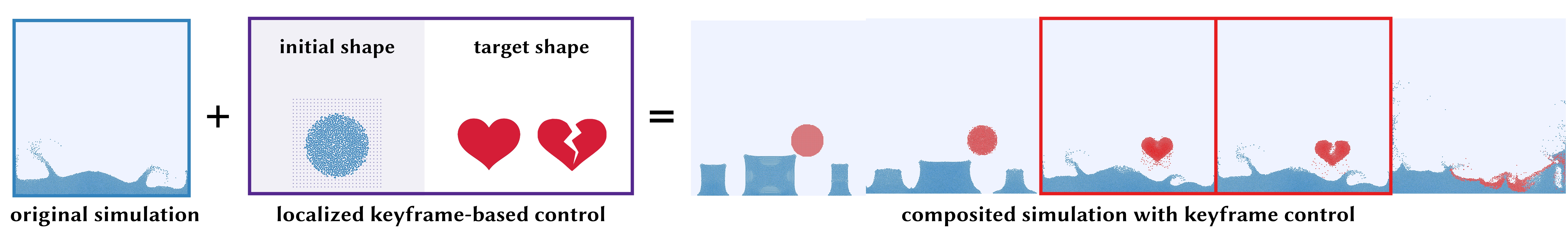}
    \caption{\textbf{2D Image-based keyframe control.} Starting from an uncontrolled simulation (left), users specify two image-based keyframes as target shapes (middle). The control forces are optimized within a localized spacetime region, producing the animation from initial shapes to desired target shapes. By compositing the optimized control forces into the global simulation (right), the fluid naturally forms the specified shapes at the correct time steps while maintaining realistic dynamics.}
    \label{fig:keyframe_based_control_2d}
\end{figure*}

\begin{figure*}[!ht]
    \centering
    \includegraphics[width=\textwidth]{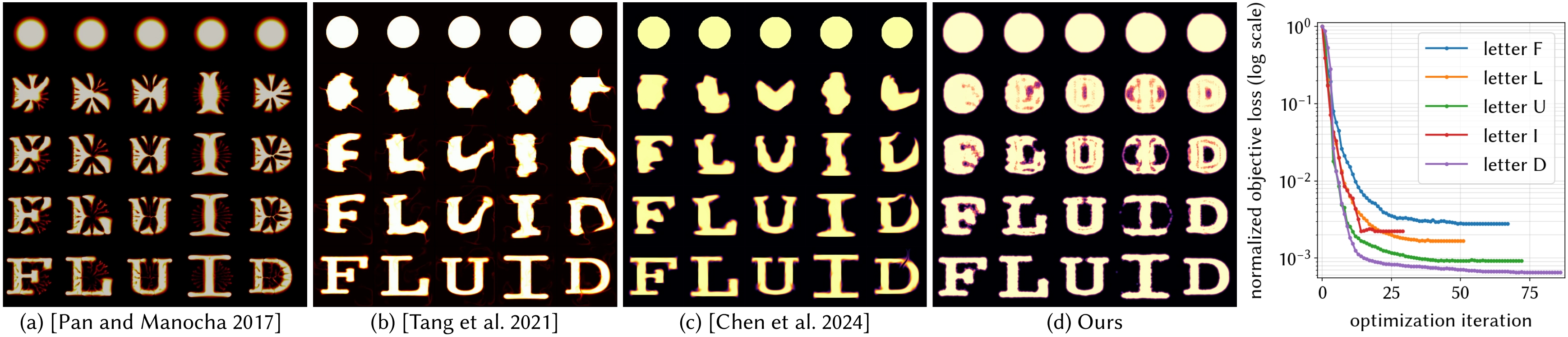}
    \caption{\textbf{Shape transformation from circles to letters F/L/U/I/D.} Given five user-specified keyframes, we transform initial circles into target letters over 40 time steps. The prior Eulerian method~\cite{pan-2017} (a) suffers from high-frequency artifacts, while~\cite{tang-2021} (b) improves performance using reduced force representations. Recent work~\cite{chen-2024} (c) employs the eigenfluid pipeline with the adjoint method to achieve smoother and faster smoke control, but remains grid-based and cannot handle free-surface flow. In contrast, our method (d) operates on free-surface particle-based liquids using a weakly compressible formulation. Despite the different simulation settings, our method operates directly on particle-based free-surface liquids while achieving comparable shape-control behavior.}
    \label{fig:2d_FLUID}
\end{figure*}

\subsubsection{Editing-Related Constraints}\label{sec:editing_constraints}
The editing-related term $\phi_{\text{editing}}$ measures how well the controlled result conforms to user-specified goals, such as desired particle transformations, user-specified pathlines, or specific splash keyframes. 

\paragraph{Particle-Based Editing Error:}
Particle-based editing can be used in two ways in our system: 1) particle-based keyframes through splash transformations (translation, scaling, and rotation), which specify the location a group of particles should be in at a specific time, and 2) pathlines, which guide specific particles to follow a user-specified trajectory forward or backward in time. With our decoupled grid-based control forces, users can define pathlines sparsely—such as only specifying the tip of a splash or a tiny part of the water volume. Due to force smoothing from our grid representation and penalization terms in~\S\ref{sec:force_constraints}, large portions of the fluid respond cohesively to minimal user input, producing consistent splash dynamics. The particle-based editing error is formulated as:
\begin{equation}
\phi_{\text{editing}}^p =
\frac{k_e}{n_p} \sum_{t \in \mathcal{K}_t}\sum_{p} w_t^p \left\| \mathbf{x}_t^p - \mathbf{x}_t^{p*} \right\|_2^2,
\end{equation}
where $\mathbf{x}_t^p$ and $\mathbf{x}_t^{p*}$ are the simulated and target positions of particle $p$ at time $t$. $n_p$ is the number of controlled particles. $w_t^p$ represents the relative importance of each particle, decreasing with its distance to the editing point according to a Gaussian kernel. $k_e$ denotes the scaling parameter of the editing error term, and $\mathcal{K}_t$ is the set of keyframes.

\paragraph{Grid-Based Editing Error:}
Keyframes can also be defined as a density distribution, by projecting particle data onto the background grid. This enables a softer, spatially distributed form of control, where constraints are not applied to individual particles. The grid-based editing error is defined as:
\begin{equation}
\phi_{\text{editing}}^g =
\frac{k_e}{n_g} \sum_{t \in \mathcal{K}_t}\sum_{g} \left\| \rho_t^g - \rho_t^{g*} \right\|^2,
\end{equation}
where $\rho_t^g$ and $\rho_t^{g*}$ denote the projected density state and density keyframe defined on the grid node $g$, and $n_g$ is the amount of localized grid nodes. Although the editing error is defined on the grid-based density field, it remains differentiable with respect to particle states through the projection operation, allowing gradients to be propagated back to particles during optimization.

\subsubsection{Force-Related Constraints}\label{sec:force_constraints}
In addition to editing-related error, we incorporate several force-related terms into the objective function to ensure that control goals are achieved with minimal control forces, which are also smooth across both space and time. 

\paragraph{Force Magnitude Regularization:}
To avoid overly large or unrealistic forces, we penalize the control force magnitude. Let $\mathbf{f}_t^{g}$ denote the control force applied at grid node $g$ at time $t$. The force magnitude term is defined as:
\begin{equation}
\phi^{\ff}_{\text{mag}} = \frac{k_f}{n_g} \sum_{t=0}^{T} \sum_{g} \left\| \mathbf{f}_t^g \right\|_2^2.
\end{equation}

\paragraph{Temporal Smoothness:}
To ensure temporal coherence, we penalize sudden changes in the control forces of the same grid node $g$ over time. Temporal smoothness leads to more continuous and realistic motion trajectories, reducing jitter or abrupt behavior in the animation. This is modeled as the squared difference of control forces between consecutive time steps:
\begin{equation}
\phi^{\ff}_{\text{temporal}} = \frac{k_t}{n_g (T-1)} \sum_{t=1}^{T} \sum_{g} \left\| \mathbf{f}_t^g - \mathbf{f}_{t-1}^g \right\|_2^2.
\end{equation}

\paragraph{Spatial Smoothness:}
We enforce spatial smoothness by penalizing the gradient of the control force in a neighboring grid region to avoid high-frequency variations in control forces across space. Let $\nabla \mathbf{f}_t^g$ denote the spatial gradient (computed via finite differences) of the control force at grid location $g$ and time $t$. The spatial regularization term is defined as:
\begin{equation}
\phi^{\ff}_{\text{spatial}} = \frac{k_{s}}{n_g} \sum_{t=0}^{T} \sum_{g} \left\| \nabla \mathbf{f}_t^g \right\|_2^2,
\end{equation}
where $k_s$ is a weighting coefficient and $T$ is the number of time steps. Note that $k_f$, $k_t$, and $k_s$ are all scaling parameters, which balance these force-related regularization terms. 

\paragraph{Total Force-Related Loss:}
In summary, the total force-related term is a combination of the above components:
\begin{align}
\phi_{\text{force}} = \phi^{\ff}_{\text{mag}} + \phi^{\ff}_{\text{temporal}} + \phi^{\ff}_{\text{spatial}}.
\end{align}
Each term is normalized by the number of grid nodes and temporal steps to ensure consistent scaling and to make the weights independent of spatial resolution and temporal discretization.

\subsubsection{Buffer Region Constraint}
\label{sec:buffer_constraint}
To decouple local regions from the rest of the simulation during control optimization, we constrain the optimization to not change state variables on the boundary of the control region. While this could be enforced in any simulation method, it is particularly easy to be captured in the PBF framework by defining a buffer region which surrounds the control volume.
During the optimization process, we simply enforce that particles in the buffer region stay as close as possible to their original trajectories from the original uncontrolled simulation. This is sufficient because state variables, such as velocity, are computed from particle positions. The constraint is formulated as:
\begin{equation}
\phi_{\text{buffer}} = \frac{k_b}{n_b} \sum_{t = 0}^T \sum_{p \in \mathcal{B}_t} \left\| \mathbf{x}_t^p - \mathbf{x}_{t, \text{orig}}^p \right\|_2^2,
\end{equation}
where $T$ is the number of timesteps where the control window is active, $\mathcal{B}_t$ represents the set of particles in the defined buffer region at time~$t$, $\mathbf{x}_{t, \text{orig}}^p$ is the uncontrolled particle position of particle p in the baseline simulations, $n_b$ is the total number of particles in the buffer region, and $k_b$ is the scaling parameter of this error term. The buffer thickness is set to be $>= 2h$, where $h$ is the PBF kernel radius (\S\ref{sec:pbf_overview}). This ensures that the buffer region fully decouples the control region from the bulk simulation, as the kernel is 0 for particle distances $> h$.

\begin{figure}[ht]
    \centering
    \includegraphics[width=\columnwidth]{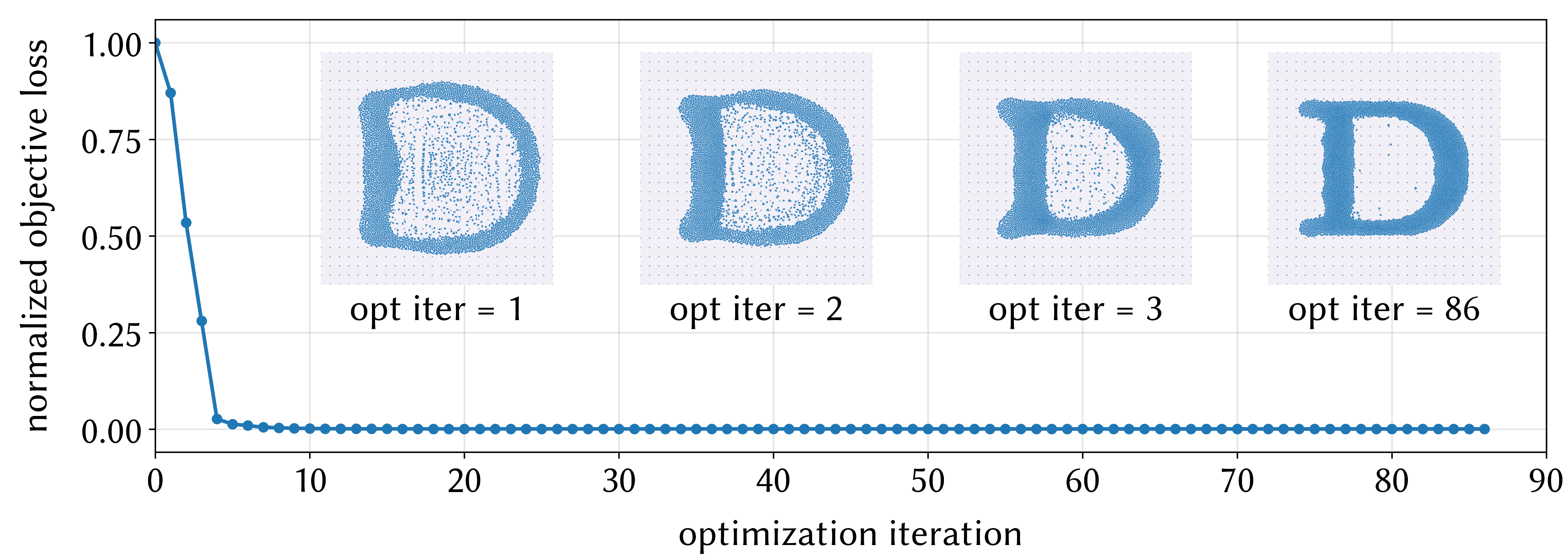}
    \caption{
    \textbf{Optimization convergence and intermediate results of letter D keyframe control.} Snapshots show the evolution of the fluid shape at selected iterations of optimization. Most perceptual improvements occur early, while later iterations mainly refine fine-scale details.
    }    
    \label{fig:letter_D_loss}
\end{figure}

\subsection{Spacetime Window Analysis and Selection}\label{sec:spacetime_window}
The effectiveness of localized fluid control depends on both the spatial extent and temporal duration of the control window. Larger windows generally provide greater control flexibility but also increase optimization complexity. Conversely, overly restrictive windows may limit the ability of control forces to influence the fluid motion required to achieve the desired edit. In this section, we first analyze the influence of spatial and temporal windows independently, then study their joint interaction through spacetime-window analysis and practical search strategies.

\begin{figure*}[!ht]
    \centering
    \includegraphics[width=\textwidth]{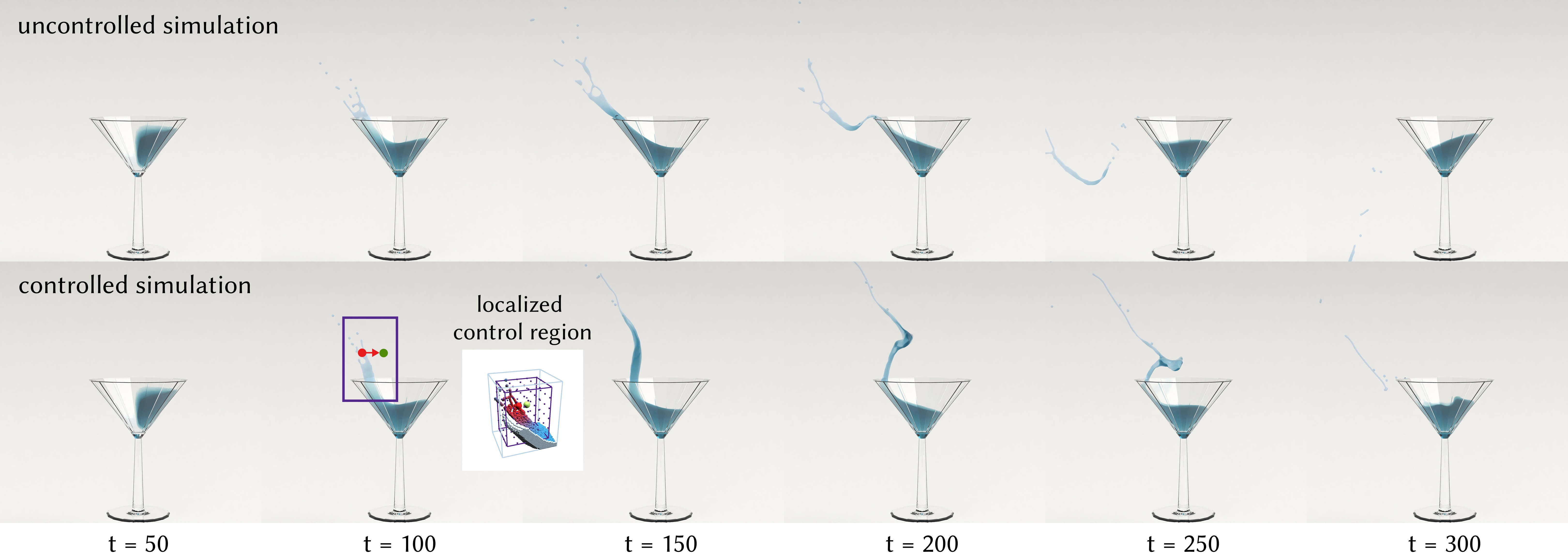}

    \vspace{0.5em}

    \small
    \setlength{\tabcolsep}{4pt}
    \renewcommand{\arraystretch}{1.1}
    \begin{tabular}{l | c | c}
    \hline
     & \# Particles in Control Region &  Backward Simulation and Gradient Computation Time (s)\\
    \hline
    Full-size Control & 78582 & 40.494 \\
    \hline
    Localized Control & 14465 & 1.985 \\
    \hline
    Reduction / Speedup & \textbf{5.4$\times$} & \textbf{20.4$\times$} \\

    \end{tabular}

    \caption{\textbf{3D martini glass fluid control.} Top: uncontrolled simulation; Bottom: our localized spacetime control. Our method restricts optimization to a bounded spacetime region (highlighted), enabling the desired control effect while significantly reducing the number of particles involved in control (78k → 14k). This reduction leads to over 20$\times$ speedup in backward simulation and gradient computation. Full-space optimization becomes impractical for simulations of this scale due to both computational and memory costs.}
    
    \label{fig:3d_splash_martini_control}
\end{figure*}

\subsubsection{Spatial Window Size}
Since the control force grid is defined locally, the size, spacing, and position of the grid have a direct impact on the effectiveness of the control. Using a full-sized high-resolution grid throughout the simulation domain would resemble traditional Eulerian control methods and largely negate the computational
advantages of localization. In contrast, applying control forces within a window that is too small can make it difficult to satisfy the control objectives due to insufficient spatial coverage.

Through our experiments (Fig.~\ref{fig:splash_editing_comparison}(a)), we observe that once the spatial control grid covers the area of interest sufficiently, the control accuracy tends to converge to a satisfactory level. Further increasing the size of the spatial window provides diminishing returns in control quality while directly increasing the control parameter DOFs (i.e., the number of the grid nodes), resulting in a more computationally expensive optimization process. 


The grid spacing within the control region introduces an additional trade-off between capturing fine-grained details and ensuring smooth transitions without high-frequency artifacts. In our experiments (Fig.~\ref{fig:splash_editing_comparison}(b)), we found that a spacing range between 10r and 20r (where r is the particle radius) works well across different scenarios. We can get efficient control feedback with a relatively coarse grid, and users can always decrease the grid spacing when more detailed or turbulent motion is desired.

\subsubsection{Temporal Window Size}
\label{sec:temporal_window_size}
The length of the temporal window over which control forces are optimized also affects the control quality, as shown in Fig.~\ref{fig:splash_editing_comparison}(c). If the temporal window is too small, the system tends to apply large, impulsive control forces over a limited duration, resulting in unrealistic motion or optimization failure. However, excessively long temporal windows significantly increase the dimensionality of the optimization problem, making it difficult to converge to an effective solution under chaotic fluid dynamics.


\subsubsection{Joint Spacetime Analysis}
Although the spatial and temporal window sizes can be analyzed independently, they are fundamentally coupled. 

To better understand this interaction, we perform a joint analysis over both spatial and temporal window sizes for the keyframe-control example (Fig.~\ref{fig:keyframe_based_control_2d}). Specifically, the spatial window size $S$ is varied between 30 and 100 grid units, while the temporal window size $T$ ranges from 5 to 35 timesteps. The lower bound of the spatial window is chosen such that the edited region is fully contained within the control domain. Fig.~\ref{fig:cma-es}(a) reports the resulting keyframe matching error, while Fig.~\ref{fig:cma-es}(b) shows the corresponding optimization complexity measured by the number of control degrees of freedom. 

Several observations can be made. First, control quality depends on both spatial and temporal extents, confirming that spacetime-window selection is inherently a joint spatial-temporal problem. Second, the objective landscape is relatively smooth over a broad range of configurations. Once a sufficiently large spatial region is selected and the temporal duration exceeds a minimum threshold, many nearby spacetime-window configurations achieve comparable control quality. Finally, although the lowest matching error is obtained using relatively large spacetime windows, these solutions also incur substantially higher optimization complexity. In practice, many near-optimal configurations achieve comparable
control quality with significantly fewer control variables.

These observations suggest that practical window selection
should focus on identifying compact spacetime regions that
provide a favorable quality-complexity trade-off rather than
precisely locating the global optimum.

\subsubsection{Practical Window Selection}
Motivated by the observations above, we first employ the Covariance Matrix Adaptation Evolution Strategy (CMA-ES)~\cite{hansen-2006} for spacetime-window selection. CMA-ES is particularly attractive in our setting because it is derivative-free, highly parallelizable across candidate evaluations, and only needs to identify a good trade-off region rather than an exactly converged global optimum. 

\begin{figure*}[ht]
    \centering
    \includegraphics[width=\textwidth]{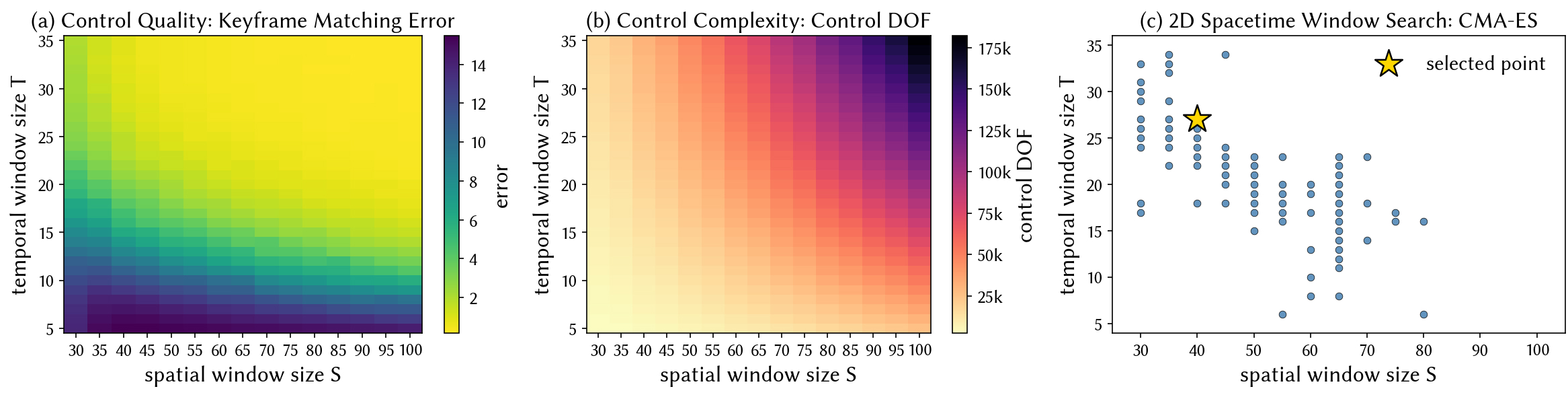}
    \caption{\textbf{Spacetime-window analysis and practical selection.} (a) The keyframe matching error of 2D keyframe control example (Fig.~\ref{fig:keyframe_based_control_2d}) for different spatial and temporal window sizes. (b) Corresponding optimization complexity measured by the number of control degrees of freedom. (c) CMA-ES search trajectory under the combined objective of control quality and optimization complexity. The selected solution lies near the low-error region while requiring substantially fewer control variables than the global optimum.}
    \label{fig:cma-es}
\end{figure*}

Fig.~\ref{fig:cma-es}(c) visualizes the 2D CMA-ES search process. Rather than minimizing the keyframe control error alone (Eq.~\ref{equ:objective_func}), the search objective balances control quality and optimization complexity by penalizing unnecessarily large control windows. More specifically, for a candidate spacetime window $W=(S,T)$, we first solve the localized control optimization problem defined in Eq.~\ref{equ:objective_func}. We then evaluate the resulting control quality together with a complexity penalty based on the number of control degrees of freedom:
\begin{equation}
\Phi(W) =
\mathcal{L}{\mathrm{control}}(f^\star(W))
+
\lambda \mathcal{C}(W),
\end{equation}
where $\mathcal{L}{\mathrm{control}}$ is the optimized control objective, $\mathcal{C}(W)$ denotes the control complexity induced by the spacetime window, and $\lambda$ balances control quality and optimization cost. Here,
C(W) is defined as the normalized number of control degrees
of freedom and we use $\lambda$ = 0.1 for this example, which gives
approximately equal importance to the normalized control objective and complexity penalty. As shown in Fig.~\ref{fig:cma-es}(c), the selected solution typically lies near the low-error region of the search space while requiring substantially fewer control degrees of freedom than the globally optimal configuration.

However, evaluating a candidate spacetime window requires solving a complete localized control problem, making exhaustive search expensive for large examples. In practical scenarios, users typically already know where the desired edit
should occur and we find the users can easily specify a sufficient spatial window in most cases (e.g., for controlling a splash, users naturally define a spatial window that covers both the formation point of the splash and the editing point at the tip of the splash). Therefore, we treat the spatial control region as user-specified according to editing intent and CMA-ES can then be used to search for an effective temporal control window size $T^\star$ within a bounded range $[T_{\min},T_{\max}]$ (see Alg.~\ref{alg:cma-es-alg}). In our experiments, the lower bound of the search range $T_{\min}$ is 5 time steps, which is chosen to prevent impulsive behavior by ensuring the control has enough time to take effect smoothly. The upper bound $T_{\max}$ is set to maintain computational efficiency and to ensure that particles remain within the influence of the spatial control window during the optimization, which is set based on the control domain size and velocity. In most examples, we set this to be 30 time steps, and start the search with $T_0$ = 10.

For applications where rapid feedback is preferred, users may alternatively determine the temporal window directly using a simple heuristic based on the spatial control window size and the particles' velocity. After specifying the spatial region of interest, particles are traced backward in time until they leave the spatial region. The temporal window is set to start at this time or 30 timesteps, whichever is smaller. In our experiments, this simple heuristic often produces results comparable to automatic search while incurring negligible additional computational cost. This observation further supports our central premise that effective control is often localized in spacetime and that practical window selection only requires identifying a reasonable control region rather than precisely optimizing its boundaries.

\begin{table*}[htb]
\label{table:timing_results}
\centering
\begin{tabular}{ccccccccccc}
\toprule
{Example} & {\thead{\# Particles \\ in Forward \\ Simulations}} & {\thead{\# Particles \\ in Control \\ ($n_p$)}} & {dt} & {\thead{Spatial \\ Window \\ Size}} &  {\thead{Grid \\ Spacing}} &  {\thead{\# Grid Nodes \\ in Control \\ ($n_g$)}} & {\thead{Temporal \\ Window \\ Size}} & {\thead{Forward \\ Simulation \\ (s/iter)}} &  {\thead{Backward \\ Simulation \\ and Gradient \\ Computation \\ (s/iter)}} & \thead{Editing \\ Error \\ Type} \\
\midrule
Fig.~\ref{fig:teaser} (a) & 1.08M & 2.4k & 0.025 & $20\times30\times15$ & 5 & 140 & 30 & 1.931 & 0.256 & particle
\\
Fig.~\ref{fig:teaser} (b) & 1.08M & 18k & 0.025 & $24\times32\times20$ & 4 & 378 & 30 & 1.931 & 2.365 & particle\\
Fig.~\ref{fig:teaser} (c) & 1.08M & 19k & 0.025 & $30\times20\times20$ & 5 & 175 & 20 & 1.931 & 3.281 & particle\\
Fig.~\ref{fig:global_vs_localized_force} (c) & 2k & 2k & 0.05 & $15\times15$ & 3 & 36 & 15 & 0.003 & 0.105 & particle
\\
Fig.~\ref{fig:keyframe_based_control_2d} (c) & 2k & 2k & 0.05 & $40\times40$ & 2 & 441 & 27 & 0.007 & 0.065 & grid
\\
Fig.~\ref{fig:2d_FLUID} (d) & 6k & 6k & 0.05 & $25\times25$ & 4 & 49 & 40 & 0.012 & 0.196 & grid
\\
Fig.~\ref{fig:3d_splash_martini_control} &  78k & 14k & 0.05 & $20\times25\times20$ & 5 & 150 & 20 & 0.461 & 1.985 & particle
\\
Fig.~\ref{fig:3d_splash_crown_control}& 338k & 17k & 0.05 & $40\times16\times20$ & 4 & 330 & 15 & 0.936 & 2.862 & particle
\\
Fig.~\ref{fig:3d_box_water_control} &  195k & 14k & 0.033 & $20\times25\times20$ & 5 & 150 & 20 & 1.112 & 1.761 & particle
\\
\bottomrule
\end{tabular}
\caption{\textbf{Execution time and configuration details for all examples.} We report the number of particles in the forward simulation and within the control region ($n_p$), time step $\Delta t$, spatial and temporal control window sizes, grid spacing, the number of grid nodes used for control ($n_g$), and the editing error type of each example. We also provide average per-iteration timings for forward simulation and backward simulation with gradient computation.}
\end{table*}

\subsubsection{Scaling Parameters}
The coefficients $k_e, k_f, k_t, k_s,$ and $k_b$ balance the contributions of the editing terms, the force regularizers and the buffer region constraint. Our objective terms are normalized by counts ($n_p, n_g, T$) to be roughly scale-invariant with respect to the number of controlled particles, grid nodes, and timesteps. Therefore, we set $k_f=k_t=k_s=k_b=1.0$ in all examples and adjust only $k_e$ according to the specific control task to balance editing accuracy against regularization.

\begin{algorithm}[t]
\caption{Practical Temporal Window Selection via CMA-ES}
\label{alg:cma-es-alg}

\KwIn{User-specified spatial control region $\Omega$ \\
      \Indp Temporal window bounds $[T_{\min}, T_{\max}]$ \\
      Initial guess $T_0$}
\KwOut{Optimal temporal window size $T^\star$}

\BlankLine
Initialize CMA-ES with mean $\mu \leftarrow T_0$ and step size $\sigma$ \;

\While{CMA-ES not converged}{
    Sample candidate window sizes $\{T\}$ from current search distribution \;

    Project to bounds and discretize:
    $T \leftarrow \mathrm{round}(\mathrm{clip}(T, T_{\min}, T_{\max}))$ \;

    \ForEach{candidate $T$}{
        Define temporal window $[t_s, t_e]$ with duration $T$ \;

        Initialize control forces $\mathbf{f} \leftarrow \mathbf{0}$ \;

        Solve the inner optimization:
        \[
        \mathbf{f}^\star = \arg\min_{\mathbf{f}} \Phi(\mathbf{f}; \Omega, [t_s, t_e])
        \]

        Evaluate objective:
        \[
        \Phi_T = \Phi(\mathbf{f}^\star; \Omega, [t_s, t_e])
        \]
    }

    Update CMA-ES using $\{T, \Phi_T\}$ \;
}

\Return{$T^\star = \arg\min_T \Phi_T$}
\end{algorithm}

\section{Results} \label{sec:results}
\begin{figure*}[!ht]
    \centering
    \includegraphics[width=\textwidth]{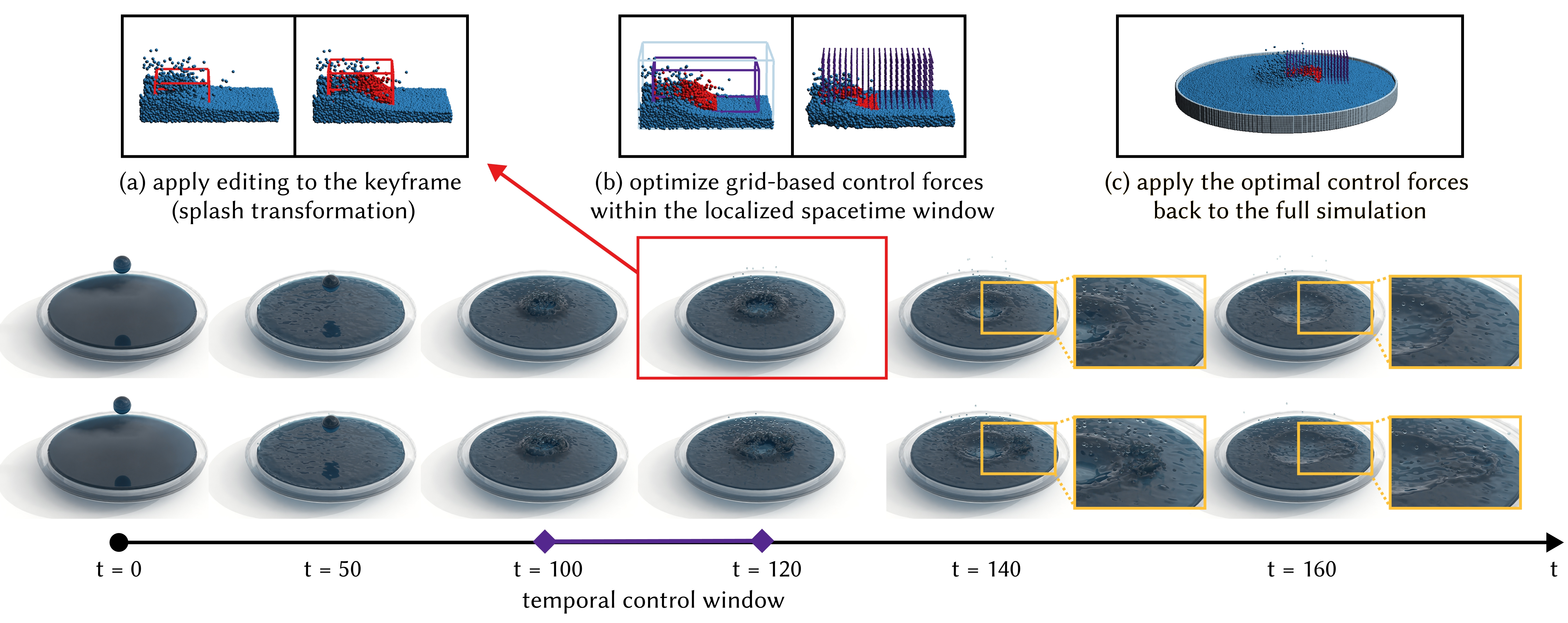}
    \caption{\textbf{3D crown splash editing control.}
   Users can apply transformations to the control box, and the splash inside will be edited accordingly. Insets highlight the differences between the original and controlled simulations, demonstrating how the edited, asymmetrical crown splash is accurately captured and seamlessly integrated into the global fluid motion.}
    \label{fig:3d_splash_crown_control}
\end{figure*}

In this section, we demonstrate our method through a series of free-surface fluid editing tasks, covering both small-scale validations and large-scale liquid scenes. All experiments are conducted on a Linux workstation equipped with an Intel Core i9 processor with 64 GB of RAM and an NVIDIA GeForce RTX 3090 GPU. 

We implement our entire algorithm in Python, leveraging the high-performance parallel programming library Taichi~\cite{hu-2019}. Taichi enables efficient GPU-accelerated and differentiable programming, significantly improving both the runtime performance and the simplicity of our implementation. For optimization, we use the L-BFGS~\cite{LBFGS} solver provided by the SciPy~\cite{SciPy-NMeth} library on CPU. We also utilize pycma~\cite{hansen-2019} for CMA-ES optimization, NumPy~\cite{harris-2020array} for general numerical operations and Matplotlib~\cite{Hunter-2007} for generating 2D plots and visualizations. To visualize our 3D results, we employ the Splashsurf~\cite{LBJB-23} library to convert the particle data (stored as PLY files from Taichi) into surface meshes in OBJ format. The final rendering results are generated using Blender~\cite{Blender}.

\begin{figure*}[!ht]
    \centering
    \includegraphics[width=\textwidth]{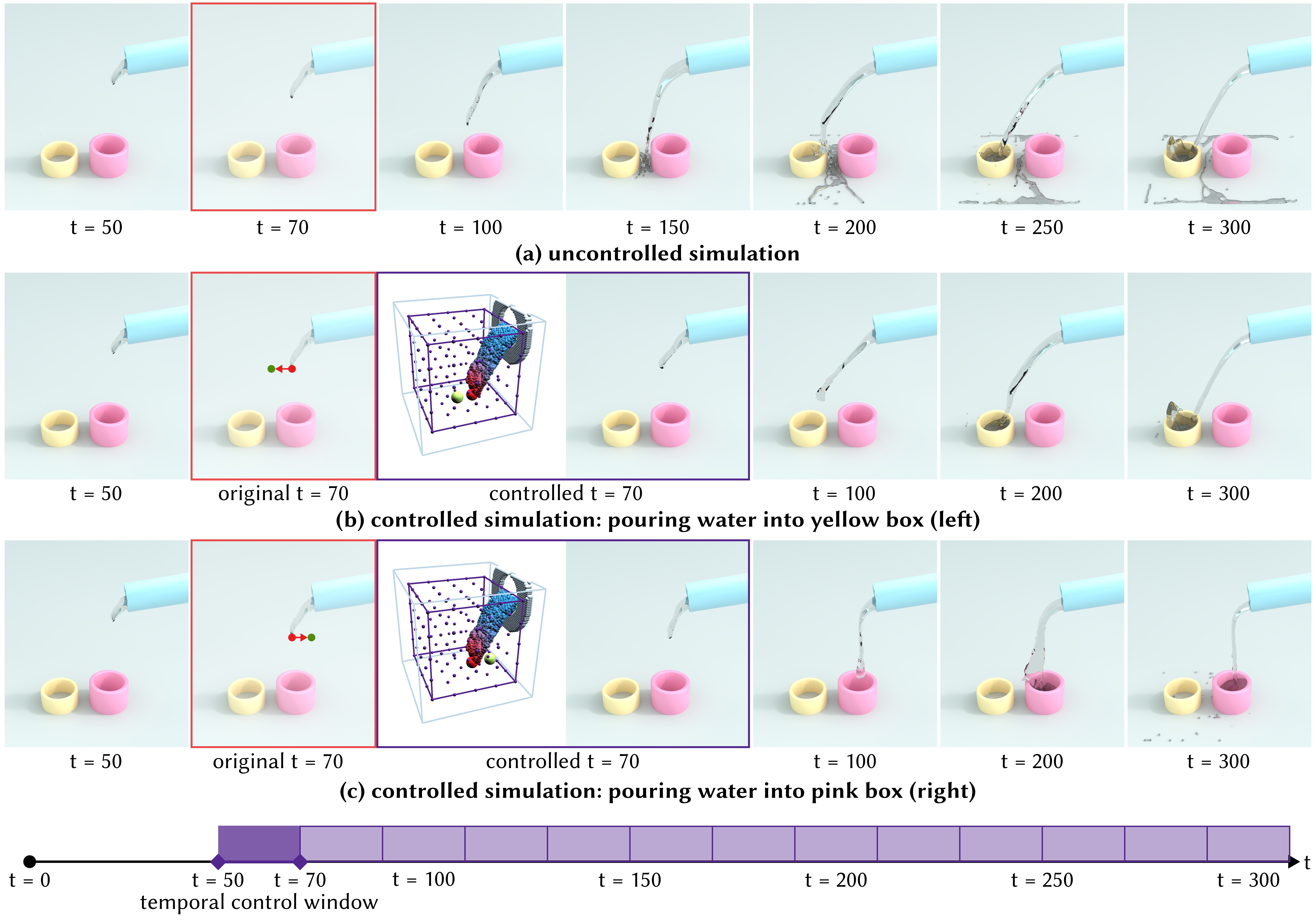}
    \caption{
    \textbf{Localized spacetime control for directing fluid into target containers.} (a) Uncontrolled simulation, where the fluid stream fails to enter either container.
    (b--c) Controlled simulations with localized edits applied at $t=70$ (highlighted in red), steering the flow into the left (yellow) or right (pink) container, respectively. Control forces are optimized within a localized spacetime window (from t=50 to t=70) (visualized by the purple grid) and applied iteratively during simulation as localized force templates.
    }
    \label{fig:3d_box_water_control}
\end{figure*}

\subsection{Performance and Scaling of Localized Spacetime Control}

A key advantage of our localized formulation is the substantial improvement in both computational cost and memory usage, particularly when the control region is small relative to the full simulation domain. As shown in Fig.~\ref{fig:teaser}, the active control region contains only a tiny fraction ($0.2\%$ - $1.7\%$) of the total particles, and the control DOFs are fewer than 400, significantly reducing the size of the optimization problem and enabling much faster convergence compared to full-space control.

Even in smaller-scale examples, where the ratio between the control region and the full domain is less extreme, our method still provides noticeable performance gains. For example, in the 3D martini glass editing example (Fig.~\ref{fig:3d_splash_martini_control}), the number of particles involved in control is reduced by $5.4\times$, resulting in more than $20\times$ speedup in backward simulation and gradient computation.

In addition, full-space optimization is highly memory-intensive and quickly becomes infeasible for large 3D simulations on a single GPU. In practice, the combination of forward simulation and backward gradient computation can exhaust GPU memory, preventing full-space optimization from running altogether. 

\subsection{Editing Modalities: Pathlines vs Keyframes}
Our framework supports several types of user interaction within a unified optimization formulation. In particular, our pipeline accommodates image-based keyframe control (Fig.~\ref{fig:keyframe_based_control_2d} and Fig.~\ref{fig:2d_FLUID}), splash keyframe editing through transformation (Fig.~\ref{fig:splash_editing_comparison} and Fig.~\ref{fig:3d_splash_crown_control}), and pathline control (Fig.~\ref{fig:global_vs_localized_force}, Fig.~\ref{fig:3d_box_water_control} and Fig.~\ref{fig:realtime_UI_2d}). For image-based keyframe control, we employ a grid-based editing objective by comparing density fields reconstructed on a background grid. In contrast, splash editing and pathline control use particle-based objectives defined directly on particle positions and trajectories. This allows the optimization objective to be adapted to the nature of the user-specified edit while preserving a common control framework.

\subsubsection{Image-based Keyframe Control} Fig.~\ref{fig:keyframe_based_control_2d} shows that localized control forces can steer the liquid toward multiple user-specified target shapes while preserving plausible surrounding dynamics. Fig.~\ref{fig:2d_FLUID} further demonstrates an image-based keyframe control example, where an initial circular configuration is driven to match a target shape (“F/L/U/I/D”) in 40 time steps. In terms of performance, the optimization of our localized control completes in approximately 4 minutes, compared to 15 minutes reported by~\cite{pan-2017}. It is important to note that the compared methods~\cite{pan-2017,tang-2021,chen-2024} focus on incompressible Eulerian smoke control, whereas our method targets free-surface particle-based liquids using a weakly compressible formulation. As a result, strict incompressibility preservation is not the primary goal of our approach. Instead, this example is intended to demonstrate that shape-driven control objectives commonly studied in smoke-control literature can also be achieved for free-surface liquids with practical optimization cost and visually coherent results.

\subsubsection{Splash Editing Control} 
Fig.~\ref{fig:3d_splash_crown_control} demonstrates that local geometric edits can be applied to a selected region of a crown splash and then seamlessly integrated back into the full simulation. In this example, a subset of splash particles is selected and transformed using simple geometric operations (e.g., scaling, translation, and rotation), and the control objective is defined via particle-to-particle matching between the transformed target and the simulated particles.

The resulting splash exhibits the desired asymmetric deformation, while the surrounding flow remains consistent with the original animation. This highlights that our control acts as a targeted local correction rather than globally overriding the simulation, enabling edited regions to blend naturally with the uncontrolled bulk motion. Similar behavior is observed in the large-scale splash scenes of Fig.~\ref{fig:teaser}, where only selected events are modified while the overall dynamics remain unchanged.

\subsubsection{Pathline Control}
\begin{figure}[!ht]
    \centering
    \includegraphics[width=\columnwidth]{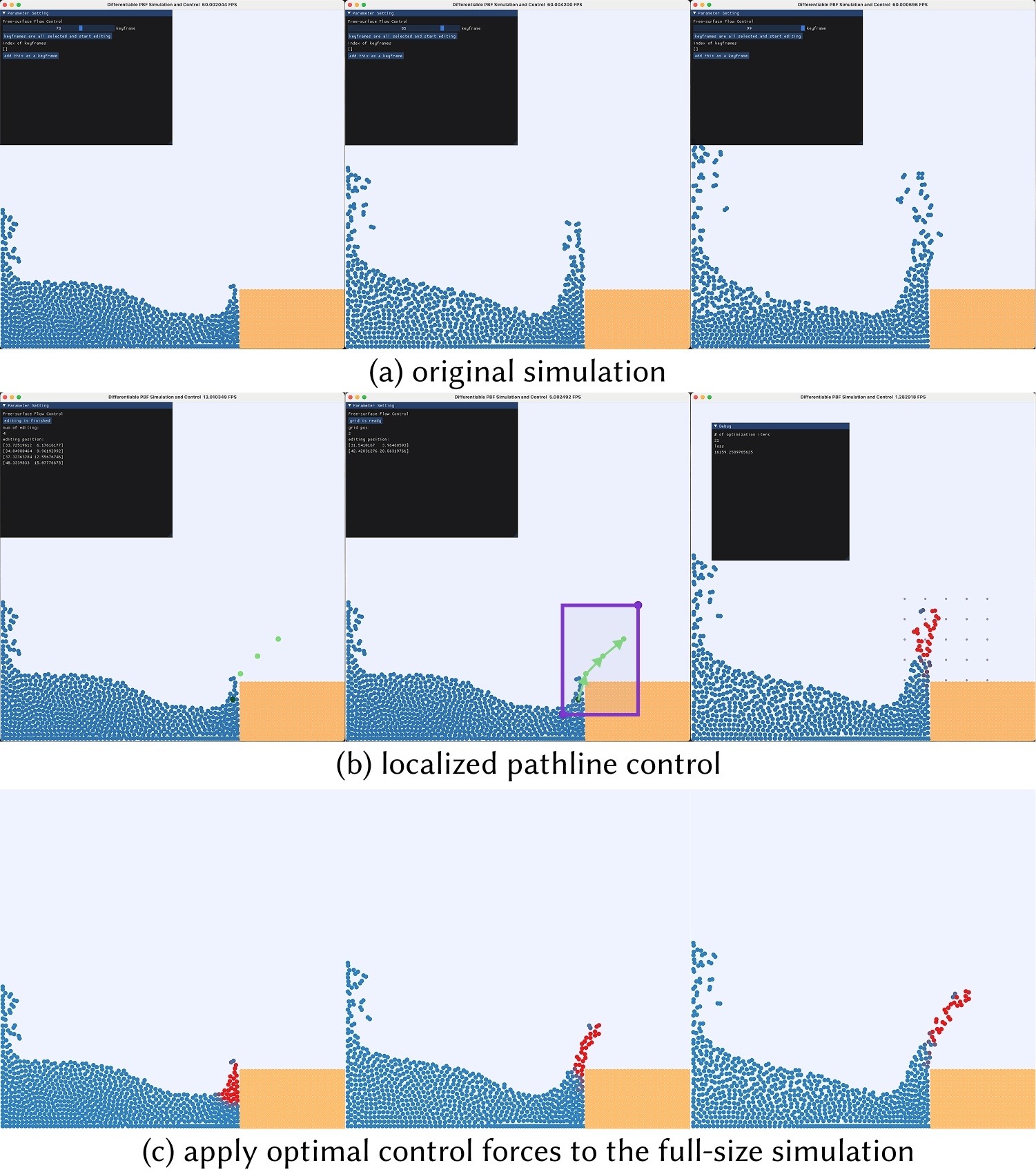}
    \caption{\textbf{Real-time 2D pathline control workflow.} The real-time workflow begins with an original fluid simulation without control (a), then applies pathline control by specifying a background control grid, target pathlines, and control parameters (b), and finally optimizes control forces on the grid and integrates them into the full simulation to steer particles along the desired trajectory (c).}
    \label{fig:realtime_UI_2d}
\end{figure}

Our method supports pathline control with sparse user inputs. In our examples (Fig.~\ref{fig:3d_splash_martini_control} and Fig.~\ref{fig:realtime_UI_2d}), pathlines are specified only near the tip of a splash and used as trajectory constraints to guide the motion of selected fluid features. Despite this localized specification, the optimized grid-based control forces induce coherent motion in the surrounding liquid, rather than affecting only a few isolated particles.

This behavior indicates that localized control signals are naturally propagated through the underlying fluid dynamics, producing spatially consistent edits without disrupting the global flow structure. Pathline control is particularly effective for directional manipulation tasks, where the goal is to steer the motion of a local fluid feature while preserving the overall flow evolution. As shown in Fig.~\ref{fig:3d_splash_martini_control}, even sparse pathline constraints are sufficient to redirect the splash in a controlled and visually plausible manner.

\begin{figure}[ht]
    \centering
    \includegraphics[trim = 40 0 0 0, clip, width=\columnwidth]{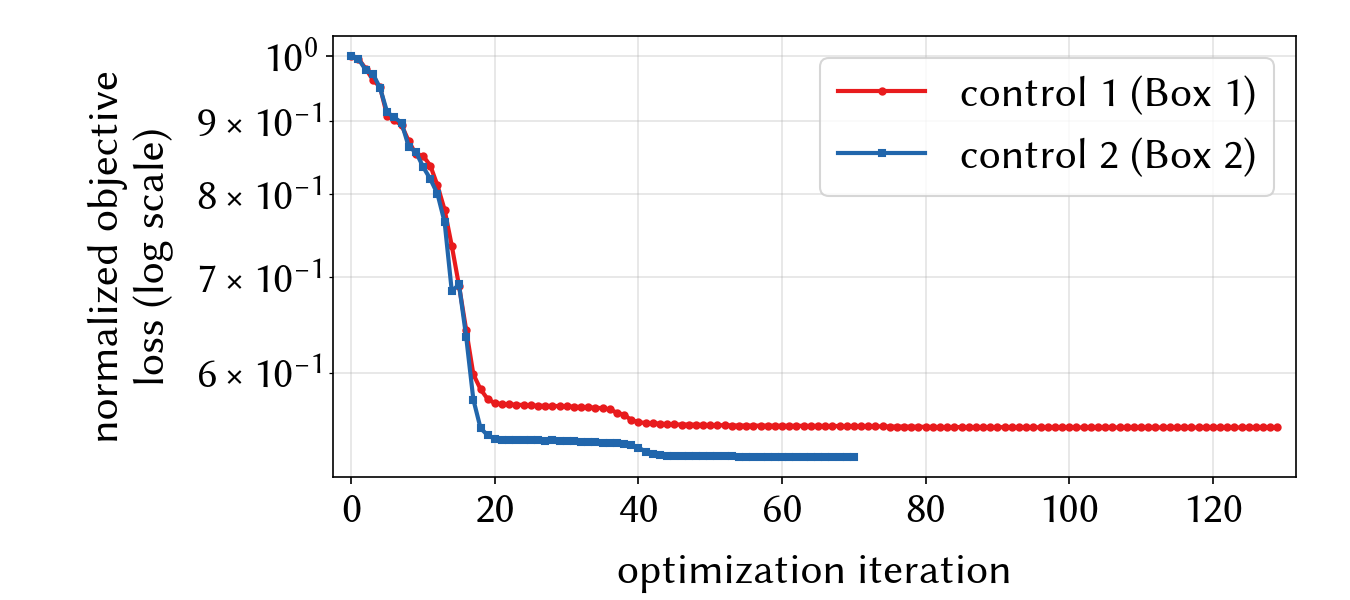}
    \caption{\textbf{Optimization convergence of localized control.} We plot the normalized objective loss over optimization iterations for the water-pouring control scenarios (Fig.~\ref{fig:3d_box_water_control}), directing the fluid stream into two different target containers. }
    \label{fig:water_pouring_loss}
\end{figure}



\subsection{Force Templates}
Interestingly, the optimized control forces within a localized temporal window can be interpreted as reusable force templates. In Fig.~\ref{fig:3d_box_water_control}, although these forces are optimized only over a short interval (from t = 50 to t = 70), replaying them throughout the simulation still produces coherent and stable redirection of the fluid stream. This behavior arises from the fact that fluid motion in certain regions exhibits repeated and structurally similar patterns. For example, in our water-pouring scenario (Fig.~\ref{fig:3d_box_water_control}), fluid particles emerging from the outlet follow similar trajectories and local configurations over time. As a result, applying the same control forces near the outlet effectively produces consistent modifications to the flow, even when reused across different time intervals. This suggests that localized optimization captures a reusable control pattern rather than a one-off correction. Despite the temporal window covering only a portion of the flow, the resulting control remains smooth and physically plausible, avoiding artifacts such as unnatural or “ghost-force” behavior.



\subsection{Convergence Behavior}
We demonstrate the convergence behavior of letter D in Fig.~\ref{fig:letter_D_loss}, where the objective decreases rapidly within the first few iterations and stabilizes after approximately 5 iterations. As shown by the intermediate snapshots, most perceptual improvements occur early in the optimization, while later iterations primarily refine fine-scale details. This fast convergence can be attributed to our localized formulation, which reduces the dimensionality of the optimization by restricting control to a bounded spacetime region. Furthermore, the grid-based force representation and smoothness regularization suppress high-frequency updates, leading to stable and well-conditioned optimization. 

We observe similar convergence behavior in the water pouring task (Fig.~\ref{fig:water_pouring_loss}), where two different control targets are considered. In both scenarios, the objective decreases rapidly during the initial iterations and reaches a plateau after approximately 20 iterations. Despite the different target containers and resulting flow directions, the convergence patterns remain consistent.
\section{Conclusion and Future Work}\label{sec:conclusion}
In this paper, we present a localized spacetime optimization framework for fluid control that exploits the observation that effective control solutions for localized editing objectives are often concentrated in bounded spacetime regions. We demonstrate its effectiveness on free-surface flows using a differentiable PBF solver. While our implementation focuses on particle-based simulation, the formulation is compatible with other differentiable solvers in principle. Though not applicable in all scenarios, when localized edits are desired, the large speedups achieved compared to full-state fluid control enable new and intuitive workflows.

There are multiple directions for future work we are excited about. Our windows are currently assumed to be non-overlapping. Allowing coupling/blending of overlapping windows could enable more precise control, e.g., for splashes off of two obstacles that are close together. Editing of control windows currently needs to be done sequentially; while windows are isolated from the bulk simulation while they are active, their changes can affect the simulation (and therefore other control windows) at later times. A way to isolate effects of windows at different times would allow a more non-linear editing workflow.

Our implementation currently only implements static objects. A differentiable handling of dynamic objects, and a corresponding update to handling control window boundary conditions, would enable more intricate examples and new modalities of control. Our implementation currently uses a non-differentiable nearest neighbor search, which causes our control timing results to be slower than necessary (as all particles in the control volume need to be searched). Replacing this with a fully differentiable and efficient alternative could further improve performance.

Finally, our results suggest that localized control solutions can be interpreted as reusable force templates. Leveraging such templates—either by reapplying them at different times within a simulation or transferring them across similar flow scenarios—represents a promising direction for improving both efficiency and user control. Fig.~\ref{fig:3d_box_water_control} provides an initial example of this capability.
\section*{Acknowledgements}

We thank all reviewers for their valuable feedback and suggestions on this work. We would also like to thank John Hancock and Xuan Dam for their administrative support. This work is supported by funding from the NSERC Discovery Grant.

\bibliographystyle{eg-alpha-doi} 
\bibliography{src/fluids}     


\end{document}